\documentclass[aps,noshowpacs,tightenlines,nobalancelastpage,floatfix,twocolumn]{revtex4}

\usepackage{amssymb}
\usepackage{amsmath}
\usepackage{graphicx}
\usepackage{dcolumn}
\usepackage{bm}

\begin{document}

\title{A One-Atom Laser in a Regime of Strong Coupling}
\date{\today }
\author{J.~McKeever, A.~Boca, A.~D.~Boozer, J.~R.~Buck, and H.~J.~Kimble}
\affiliation{Norman Bridge Laboratory of Physics 12-33, California Institute of
Technology, Pasadena, CA 91125}
\maketitle

\textbf{Although conventional lasers operate with a large number
of intracavity atoms, the lasing properties of a single atom in a
resonant cavity have been theoretically investigated for more than
a decade
\cite{mu92,ginzel93,pellizzari94a,pellizzari94b,horak95,meyer97a,loeffler97,meyer97b,meyer98,jones99,kilin02}.
In this Letter we report the experimental realization of such a
one-atom laser operated in a regime of strong coupling. Our
experiment exploits recent advances in cavity quantum
electrodynamics that allow one atom to be isolated in an optical
cavity in a regime for which one photon is sufficient to saturate
the atomic transition \cite{mckeever03}. In this regime the
observed characteristics of the atom-cavity system are
qualitatively different from those of the familiar many atom case.
Specifically, we present measurements of intracavity photon number
versus pump intensity that exhibit \textquotedblleft
thresholdless\textquotedblright\ behavior, and infer that the
output flux from the cavity mode exceeds that from atomic
fluorescence by more than tenfold. Observations of the
second-order intensity correlation function} $g^{(2)}(\tau)$
\textbf{demonstrate that our one-atom laser generates manifestly
quantum (i.e., nonclassical) light that exhibits both photon
antibunching} $g^{(2)}(0)<g^{(2)}(\tau)$ \textbf{and
sub-Poissonian photon statistics} $g^{(2)}(0)<1$\textbf{.}

An important trend in modern science is to push macroscopic
physical systems to ever smaller sizes, eventually into the
microscopic realm. Lasers are one important example of this
progression, having moved from table-top systems to microscopic
devices that are ubiquitous in science and technology. However,
even over this remarkable span of implementations, lasers are
typically realized with large atom and photon numbers in a domain
of \textit{weak coupling} for which individual quanta have
negligible impact on the system dynamics. Usual laser theories
therefore rely on system-size expansions in inverse powers of
critical atom and photon numbers $(N_{0},n_{0})\gg 1$, and arrive
at a consistent form for the
laser characteristics \cite%
{sargent-book,haken-book,mandel-wolf-95,carmichael-book,gardiner-book}.
By contrast, over the past twenty years, technical advances on
various fronts have pushed laser operation to regimes of ever
smaller atom and photon number, pressing toward the limit of
\textit{strong coupling} for which $(N_{0},n_{0})\ll 1$ \cite{hjk
sweden}. Significant milestones include the realization of one and
two-photon micromasers
\cite{walther-review,haroche-review,meystre92}, as well as novel
microlasers in atomic and condensed matter systems
\cite{feld-ml,chang-campillo96,vahala03}.

In this march toward ever smaller systems, an intriguing
possibility is that a laser might be obtained with a single atom
in an optical cavity, as was considered in the seminal work of Mu
and Savage \cite{mu92} and has since been extensively analyzed
\cite{ginzel93,pellizzari94a,pellizzari94b,horak95,meyer97a,loeffler97,meyer97b,meyer98,jones99,kilin02}.
In this Letter we report an experiment that advances this quest to
its conceptual limit, namely the operation of a one-atom laser in
a regime for which $(N_{0},n_{0})\ll 1$. We describe measurements
of the operating characteristics of the atom-cavity system,
including the light output as a function of pumping strength.
Significantly, the light emission is observed to exhibit photon
antibunching and is manifestly quantum in character, in contrast
to the light from conventional lasers, thereby enabling diverse
applications in quantum optics and quantum information science.

\begin{figure}[tb]
\includegraphics[width=8.6cm]{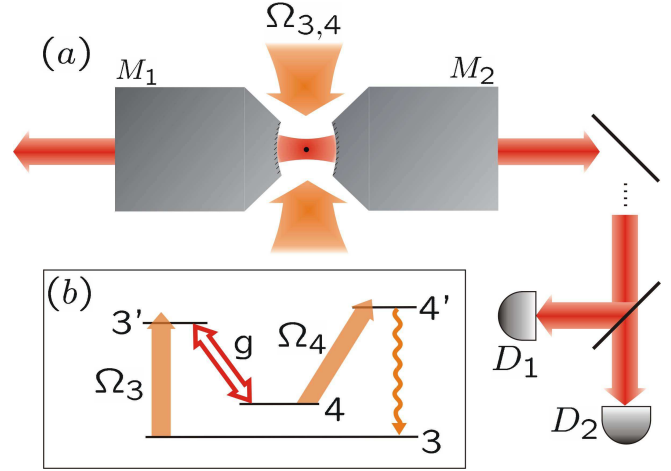}
\caption{A simplified schematic of the experiment. (a) A Cesium
atom (black dot) is trapped inside a high-finesse optical cavity
formed by the curved, reflective surfaces of mirrors
\textit{M}$_{1,2}$. Light generated by the atom's interaction with
the resonant cavity mode propagates as a Gaussian beam to
single-photon detectors $D_{1,2}$. (b) The relevant transitions
involve the $6S_{1/2},F=3,4\leftrightarrow 6P_{3/2},F^{\prime
}=3^{\prime },4^{\prime }$ levels of the $D_{2}$ line at $852.4$
nm in atomic Cesium. Strong coupling at rate $g$ is achieved for
the lasing transition $F^{\prime }=3^{\prime }\rightarrow F=4$
near a cavity resonance. Pumping of the upper level
$F^{\prime}=3^{\prime }$ is provided by the field $\Omega _{3}$,
while recycling of the lower level $F=4$ is achieved by way of the field $%
\Omega _{4}$ ($4\rightarrow 4^{\prime }$) and spontaneous decay
back to $F=3$. Decay $(3^{\prime}, 4^{\prime }) \rightarrow (3,4)
$ is also included in our model. Relevant cavity parameters are
length $l_0=42.2~\mathrm{\protect\mu m}$, waist
$w_{0}=23.6~\mathrm{\mu m}$, and finesse $\mathcal{F}=4.2\times
10^{5}$ at $\lambda_{D_{2}}=852~\mathrm{nm}$.} \label{oal-setup}
\end{figure}

As illustrated in Fig. \ref{oal-setup}, our experiment consists of
a single Cesium atom trapped in a far-off-resonance trap (FORT)
within a high-finesse optical cavity \cite{ye99,mckeever03}. The
lasing transition $6P_{3/2},F^{\prime}=3^{\prime}\rightarrow
6S_{1/2},F=4$ is nearly resonant with and strongly coupled to a
single mode of this cavity. The coupling is parameterized by the
Rabi frequency $2g_{0}$ for a single quantum of excitation, and
the atom and field have amplitude decay rates $\gamma$ and
$\kappa$, respectively. The upper level $F^{\prime }=3^{\prime }$
is pumped by the external drive $\Omega _{3}$, while effective
decay of the lower level $F=4$ takes place via the combination of
the drive $\Omega _{4}$ and decay $\gamma _{34}$, $4\rightarrow
4^{\prime }\rightarrow 3$. In essential character this system is
analogous to a Raman scheme with pumping $3\rightarrow 3^{\prime
}$, lasing $3^{\prime }\rightarrow 4$, and decay $4\rightarrow 3$.
Of particular relevance to our work are the detailed treatments of
the ion-trap laser by Walther and colleagues \cite
{meyer97a,loeffler97,meyer97b,meyer98}.

We emphasize that a \textquotedblleft
one-and-the-same\textquotedblright\ atom laser as illustrated in
Fig. \ref {oal-setup} is quite distinct from \textquotedblleft
single-atom\textquotedblright\ micro-masers
\cite{walther-review,haroche-review,meystre92} and lasers
\cite{feld-ml} for which steady state is reached through the
incremental contributions of many atoms that transit the cavity,
even if one by one \cite{walther-review,haroche-review} or few by
few \cite{feld-ml}. By contrast, in our experiment steady state is
reached with one-and-the-same atom over a time interval $\delta
t\sim 10^{-7}~\mathrm{s}$ that is much shorter than the trap
lifetime $\Delta t\sim 0.05~\mathrm{s}$. Our pumped atom-cavity
system provides a continuous source of nonclassical light as a
Gaussian beam for the entire duration that an atom is trapped.

Because conventional lasers operate in the limit $(N_{0},n_{0})\gg
1$, there is a generic form associated with the laser threshold in
the transition from nonlasing to lasing action that is independent
of the model system \cite {sargent-book,rice94}. However, as the
system size is reduced, the sharpness of the laser
\textquotedblleft turn on\textquotedblright\ is lost, with then no
clear consensus about how to define the lasing threshold
\cite{rice94}. Well into the regime of strong coupling
$(N_{0},n_{0})\ll 1$, even the familiar qualitative
characteristics of a laser (e.g., the statistical properties of
the output light) are profoundly altered, leaving open the
question of how to recognize a laser in this new regime.

To address this question, we have carried out extensive
theoretical analyses for a four-state model based upon Fig.
\ref{oal-setup}(b) for parameters relevant to our experiment. A
synopsis of relevant results from this work is given in the
\textit{Supplementary Information}, with the full treatment
presented in Ref. \cite{boozer03}. In brief, the steady state
solutions obtained from a semiclassical theory exhibit familiar
characteristics of conventional lasers, including a clearly
defined laser threshold and population inversion. The condition
$C_{1}\gg 1$ is required to observe threshold behavior for one
atom pumped inside the resonator, where for our experiment the
cooperativity parameter $C_1=1/N_{0}\simeq 12$. By contrast, the
fully quantum analysis for the four-state model results in
qualitatively different characteristics. In particular, the
input-output relationship for the mean intracavity photon number
$\bar{n}$ versus the pump intensity $I_{3}=(\Omega
_{3}/2\protect\gamma )^{2}$ has several key features to be
compared with experimental results presented below, namely the
immediate onset of emission (``thresholdless'' behavior), and the
saturation and eventual quenching of the output.

Our actual experiment is somewhat more complex than indicated by
the simple drawing in Fig. \ref{oal-setup}, with many of the
technical aspects described in more detail in Refs.
\cite{mckeever03,ye99}. In brief, the principal cavity QED (cQED)
parameters of our system are $g_{0}/2\pi =16~\mathrm{MHz}$,
$\kappa /2\pi =4.2~\mathrm{MHz}$, and $\gamma /2\pi
=2.6~\mathrm{MHz}$, where $g_{0}$ is based upon the reduced dipole
moment for the $6S_{1/2},F=4\leftrightarrow 6P_{3/2},F^{\prime
}=3^{\prime }$ transition in atomic Cs. Strong coupling is thereby
achieved $(g_{0}\gg (\kappa ,\gamma ))$, resulting in critical
photon and atom numbers $n_{0}\equiv \gamma
^{2}/(2g_{0}^{2})\simeq 0.013$, $N_{0}\equiv 2\kappa \gamma
/g_{0}^{2}\simeq 0.084$.

Atoms are trapped in the cavity by means of a far-off-resonance
trap (FORT) \cite{metcalf99} with wavelength
$\lambda_{F}=935.6~\mathrm{nm}$, which is matched to a
$\mathrm{TEM_{00}}$ mode along the cavity axis. For all
experiments herein, the trap depth is
$U_{0}/k_{B}=2.3~\mathrm{mK}$ ($47~\mathrm{MHz}$). The FORT has
the important feature that the potential for the atomic
center-of-mass motion is only weakly dependent on the atom's
internal state \cite{mckeever03}.

\begin{figure}[tb]
\includegraphics[width=8.6cm]{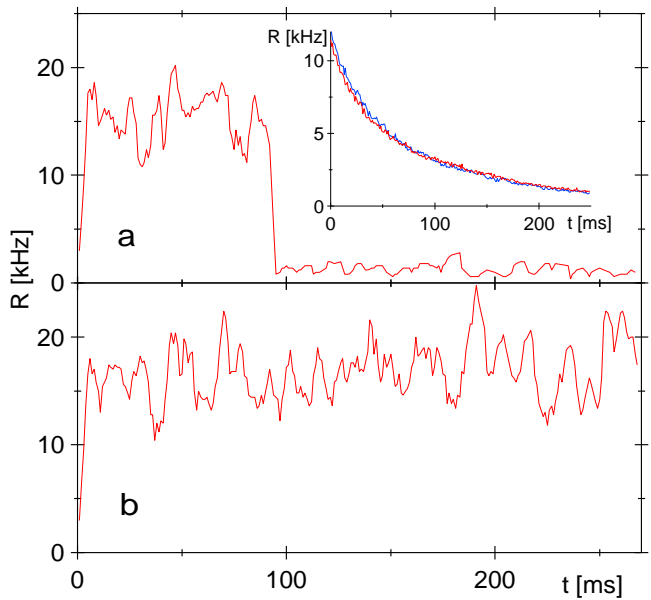}
\caption{Total counting rate $R$ recorded by detectors
\textit{D}$_{1,2}$ is displayed as a function of time for two
separate trapped atoms, with the counts summed over $5$ ms bins.
At $t=0$, the $\Omega _{3,4}$ fields are switched to predetermined
values of intensity and detuning. In (a), the atom is trapped for
$t\simeq 90$ ms before escaping, with the background level due to
scattered light from the $\Omega _{3,4}$ fields and detector dark
counts evident as the residual output at later times. In (b), the
atom (atypically) remains trapped for the entire observation cycle
$\simeq 270$ ms and then is dumped. The inset in (a) displays $R$
versus time obtained by averaging about $400$ such traces. Two
cases are shown; in one, the number of atoms delivered to the
cavity mode has been diminished by about $2$-fold. Since the
curves are nearly identical, we conclude that cases with $N>1$
atom play a negligible role. The overall detection efficiency $\xi
=0.05$ from intracavity photon to a detection event at
\textit{D}$_{1,2}$ is made up of the following factors: $\eta
=0.60$ cavity escape efficiency, $T=0.50$ for only mirror
\textit{M}$_{2}$ output, $\zeta =0.33$ propagation efficiency from
\textit{M}$_{2}$ to \textit{D}$_{1,2}$, $\alpha =0.5$ detection
quantum efficiency at \textit{D}$_{1,2}$.} \label{rvst}
\end{figure}

After the trap-loading stage (as described in the section on
\textbf{Methods}), the transverse $\Omega _{3,4}$ fields are
switched to pump and recycle the atomic population in the fashion
depicted in Fig. \ref{oal-setup}(b). Two examples of the resulting
output counts versus time are shown in Fig. \ref{rvst}. By
averaging traces such as these, we arrive at an average signal
level versus time, as shown in the inset to Fig. \ref{rvst}(a).
Typical lifetimes for a trapped atom in the presence of the
driving $\Omega _{3,4}$ fields are $50-100$ ms, which should be
compared to the lifetimes of $2-3$ s recorded in the absence of
these fields \cite{mckeever03}. Significantly, the approximately
exponential decay of the signal with time does not result from a
time-dependent diminution of the flux from single trapped atoms,
but rather from the average of many events each of a variable
duration. That is, for a given set of external control parameters,
each atom gives a reasonably well-defined output flux over the
time that it is trapped.

\begin{figure}[tb]
\includegraphics[width=8.6cm]{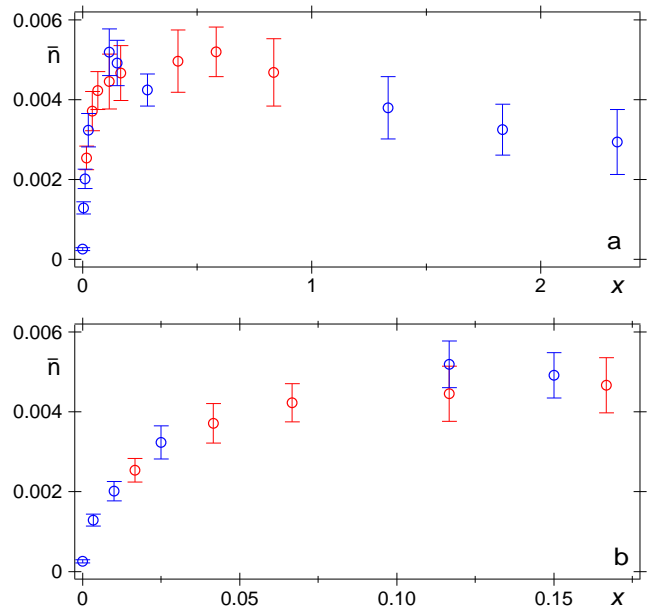}
\caption{The intracavity photon number $\bar{n}\pm
\protect\sigma_{n}$ inferred from measurements as in Fig.
\protect\ref{rvst} is plotted as a function of dimensionless pump
intensity $x\equiv (7/9)(I_3/I_4)$ for fixed $I_{4}=13$ over two
ranges of pump level $x$. (a) $\bar{n}$ versus $x$ is shown over
the entire range $x=0$ to $2.33$ recorded in our measurements (b)
An expanded scale displays $\bar{n}$ for small $x$. The immediate
onset of emission supports the conclusion of \textquotedblleft
thresholdless\textquotedblright\ lasing. The two independent sets
of measurements (red and blue points) agree reasonably well.}
\label{nvsx}
\end{figure}

For a fixed set of operating conditions, we collect a set of
60-300 traces as in Fig. \ref{rvst}, determine the average output
flux for each trace, and find the mean and variance, as well as
the trap lifetime for the set.  Figure \ref{nvsx} displays a
collection of such measurements for the mean intracavity photon
number $\bar{n}$ as a function of the dimensionless pump intensity
$x$, scaled in units of the fixed recycling intensity (see section
on \textbf{Methods}). More precisely, the parameter $x$ is the
ratio of measured intensities, and can be written as $x\equiv
(7/9)(I_{3}/I_{4})$, where $I_{3,4}\equiv (\Omega _{3,4}/2\gamma
)^{2}$. The factor of $(7/9)$ is needed because the the two
transitions have different dipole moments. For these measurements,
we estimate that the incoherent sum of intensities of the four
$\Omega_{4}$ beams is about $50~\mathrm{mW/cm^{2}}$, which
corresponds to $I_4 \sim 13$. The output count rate at detectors
\textit{D}$_{1,2}$ is converted to intracavity photon number using
the known propagation and detection efficiency $\xi =0.05$.

Important features of the data shown in Fig. \ref{nvsx} include
the prompt onset of output flux $\kappa \bar{n}$ emerging through
the cavity mirrors $M_{1,2}$ as the pump intensity $I_{3}$ is
increased from zero. In a regime of strong coupling, the
atom-cavity system behaves as a \textquotedblleft
thresholdless\textquotedblright\ device. With further increases in
pump intensity $I_{3}$, the output flux saturates at a maximum
value $\kappa \bar{n}_{\max }$ around $x\simeq 0.1$. We attribute
this behavior to a bottleneck associated with the recycling of
population $4\rightarrow 4^{\prime }\rightarrow 3$, with the rate
limiting step in the recycling process being spontaneous decay
$4^{\prime }\rightarrow 3$ at rate $\gamma _{34}$ in the limit of
large Rabi frequency $\Omega _{4}\gg \gamma $. For a single
intracavity atom, quanta can be deposited into the cavity mode no
faster than the maximum recycling rate. As the pump level $I_{3}$
is increased beyond $x\sim 1$, the output flux $\kappa \bar{n}$
gradually drops, presumably due to splitting of the pumped excited
state $F^{\prime}=3^{\prime} $ by the Autler-Townes effect,
although this is still under investigation. Heating of the atomic
motion at higher pump levels is certainly a concern as well;
however our simulations, which do not incorporate atomic motion,
show the same trend as in Fig. \ref{nvsx} \cite{boozer03}.

Beyond these considerations, we have also undertaken extensive
theoretical analyses based both upon the four-state model shown in
Fig. \ref{oal-setup}, as well as on the full set of Zeeman states
for each of the levels $F=3,4$ and $F^{\prime }=3^{\prime
},4^{\prime }$ and two cavity modes, one for each of two
orthogonal polarizations \cite{boozer03}. These analyses are in
reasonable accord with the principal features of the data in Fig.
\ref{nvsx}. Moreover, our quantum simulations support the
conclusion that the range of coupling values $g$ that contribute
to our results is restricted roughly to $0.5g_{0}\lesssim
g\lesssim g_{0}$. Furthermore, the simulations yield information
about the atomic populations, from which we deduce that the rate
of emission from the cavity $\kappa\bar{n}$ exceeds that by way of
fluorescent decay $3^{\prime }\rightarrow 4$, $\gamma _{4
3^{\prime }}\langle \sigma _{3^{\prime }3^{\prime }}\rangle $, by
roughly tenfold over the range of pump intensity $I_{3}$ shown in
Fig. \ref{nvsx}, where $\langle \sigma _{3^{\prime }3^{\prime
}}\rangle $ is the steady state population in level $3^{\prime }$.

To investigate the quantum statistical characteristics of the
light emerging in the TEM$_{00}$ mode of the cavity output, we
probe the photon statistics of the light by way of the two
single-photon detectors \textit{D}$_{1,2}$ illustrated in Fig.
\ref{oal-setup}. From the cross-correlation of the resulting
binned photon arrival times and the mean counting rates of the
signals and the background, we construct the normalized intensity
correlation function (see the \textit{Supplementary Information})
\begin{equation}
g^{(2)}(\tau )=\frac{\langle :\hat{I}(t)\hat{I}(t+\tau ):\rangle }{\langle :%
\hat{I}(t):\rangle ^{2}}\text{ ,}  \label{g2}
\end{equation}%
where the colons denote normal and time ordering for the intensity
operators $\hat{I}$ \cite{mandel-wolf-95}. Over the duration of
the trapping events, we find no evidence that $\langle
:\hat{I}(t):\rangle $ is a function of $t$, although we do not
have sufficient data to confirm quantitatively stationarity of the
underlying processes.

\begin{figure}[tb]
\includegraphics[width=8.6cm]{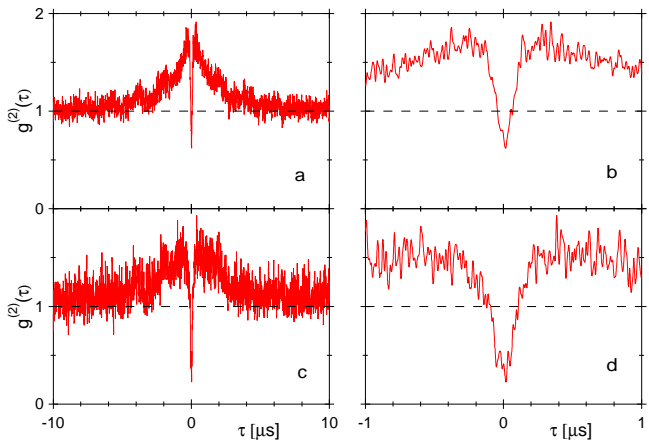}
\caption{The intensity correlation function
$g^{(2)}(\protect\tau)$ is given for two values of the pump
intensity in (a-d), where (b,d) show the central features of (a,c)
over a smaller range of $\tau$. Panels (a,b) are for pump
intensity parameter $x=0.83$, whereas for (c,d) $x=0.17$. Note
that the light exhibits sub-Poissonian photon statistics and
antibunching.  All traces have been ``smoothed'' by convolution
with a Gaussian function of width $\sigma=5~\mathrm{ns}$.}
 \label{g2tau}
\end{figure}

Examples of two measurements for $g^{(2)}(\tau )$ are given in
Fig. \ref{g2tau}(a-d). In Fig. \ref{g2tau}(a,b), we again have
$I_4\simeq 13$ and the pump intensity $I_{3}$ is set for operation
with $x \simeq 0.83$ well beyond the \textquotedblleft
knee\textquotedblright\ in $\bar{n}$ versus $x$, while in (c,d),
the pump level is decreased to $x\simeq 0.17$ near the peak in
$\bar{n}$. Significantly, in each case these measurements
demonstrate that the light from the atom-cavity system is
manifestly quantum (i.e., nonclassical) and exhibits photon
antibunching $g^{(2)}(0)<g^{(2)}(\tau )$\ and sub-Poissonian
photon statistics $ g^{(2)}(0)<1$ \cite{mandel-wolf-95}. The
actual coincidence data $n (\tau )$ used to obtain $g^{(2)}(\tau
)$ are presented in the \textit{Supplementary Information}.
Significantly, these data directly evidence the nonclassical
character of the emitted light, with relatively minor corrections
for background light required for the determination of
$g^{(2)}(\tau )$.

Beyond the nonclassical features around $\tau \simeq 0$,
$g^{(2)}(\tau )$ also exhibits excess fluctuations extending over
$\tau \simeq \pm 1~\mathrm{\mu s}$, with $g_{\max }^{(2)}(\tau
)\simeq 1.7$. Fluctuations in the intensity of the intracavity
light over these time scales are presumably related to the
stochastic character of the pumping $3\rightarrow 3^{\prime }$ and
recycling $4\rightarrow 4^{\prime }\rightarrow 3$ processes for a
single, multi-state atom. Also of significance is the interplay of
atomic motion and optical pumping into dark states by the $\Omega
_{3,4}$ fields (which is responsible for cooling \cite{boiron96}),
as well as Larmor precession that arises from residual ellipticity
in polarization of the intracavity FORT
\cite{mckeever03,corwin99}. Indeed, in Fig. \ref{g2tau}(a,c) there
is a hint of an oscillatory variation in $g_{\max }^{(2)}(\tau )$
with period $\tau \simeq \pm 2~\mathrm{\mu s}$. Fourier
transformation of the associated coincidence data leads to a small
peak at about $500$ kHz, which is near to the predicted frequency
for axial motion of a trapped Cs atom at the bottom of the FORT
potential as well as to the Larmor frequency inferred from other
measurements.

In agreement with the trend predicted by the four-state model
discussed in the \textit{Supplementary Information}, $g^{(2)}(0)$
increases with increasing pump intensity, with a concomitant
decrease in these nonclassical effects. Moreover, our experimental
observations of $g^{(2)}(\tau )$ are described reasonably well by
the results obtained from more detailed quantum simulations based
upon the entire manifold of Zeeman states for the Cs atom, two
cavity modes with orthogonal polarizations, and a simple model to
describe the polarization gradients of the $\Omega_{3,4}$ fields
\cite{boozer03}.

The realization of this strongly coupled one-atom laser is
significant on several fronts. From the perspective of the
dynamics of open quantum systems, our system demonstrates the
radical departures from conventional laser operation wrought by
strong coupling for the quantized light-matter interaction. On a
more practical level, throughout the interval when an atom is
trapped (which is determined in real time), our system provides an
approximately stationary source of nonclassical light in a
collimated, Gaussian beam, as has been anticipated in the
literature on one-atom lasers
\cite{mu92,pellizzari94a,pellizzari94b,horak95,meyer97a,meyer97b,meyer98,jones99,kilin02},
and which has diverse applications. Remaining technical issues in
our work are to improve the modelling and measurements related to
atomic motion, both within the FORT potential and through the
polarization gradients of the $\Omega _{3,4}$ fields. We have
employed our quantum simulations to calculate the optical spectrum
of the light output, and have devised a scheme for its
measurement.

\textbf{Methods} - While the atom is trapped in a standing wave
FORT along the cavity axis, another set of fields (designated by
$\Omega _{3,4}$ in Fig. \ref{oal-setup}) propagate in the plane
transverse to the cavity axis and illuminate the region between
the cavity mirrors. These fields are used not only for the pumping
scheme described in association with the operation of the one-atom
laser with strong coupling, but also for cooling in the
trap-loading phase. Each $\Omega _{3,4}$ field consists of two
orthogonal pairs of counter-propagating beams in a $\sigma
^{+}-\sigma ^{-}$ configuration.

Unfortunately, it is difficult to calibrate accurately the
intensities $I_{3,4}$ for the $\Omega _{3,4}$ beams at the
location of the atom in the region between the cavity mirrors. We
estimate that our knowledge of either intensity is uncertain by an
overall scale factor $\simeq $ $2$. However, we do know the ratio
of intensities much more accurately than either intensity
individually, and therefore plot the data in Fig. \ref{nvsx} as a
function of this ratio.

In the pumping stage of the experiment, the fields are tuned $10$
MHz blue of $F=3\rightarrow F^{\prime }=3^{\prime }$ in the case
of the $\Omega _{3}$ beams and $17$ MHz blue of $F=4\rightarrow
F^{\prime }=4^{\prime }$ in the case of the $\Omega _{4}$ fields.
The detuning between the $ 3^{\prime }\rightarrow 4$ transition at
$\omega _{43}$ and the cavity resonance $\omega _{C}$ is $\Delta
_{CA}\equiv \omega _{C}-\omega _{43}=2\pi \cdot 9$ MHz. These
detunings are chosen operationally in a trade-off between
achieving a large cavity output flux from the $3^{\prime
}\rightarrow 4$ transition while maintaining a reasonable lifetime
for the trapped atom despite heating from the various fields
\cite{boiron96}. The cavity length itself is actively stabilized
with an auxiliary laser at wavelength $\lambda
_{C}=835.8~\mathrm{nm}$ that does not interfere with the trapping
or the cQED interactions.

Our experimental protocol begins with the formation of a
magneto-optical trap (MOT) above the cavity. After a stage of
sub-Doppler cooling, the cloud of atoms is released. The
$\Omega_{3,4}$ beams are then used as cooling beams (with
independent settings of intensity and detuning) to load an atom
into the FORT \cite{mckeever03}. About 10 atoms transit the cavity
mode after each MOT drop, and the loading efficiency is set such
that an atom is loaded into the FORT once every 3-10 drops. We
then switch the intensities and detunings of the transverse fields
$\Omega_{3,4}$ to the pumping configuration and record the cavity
output by way of the single-photon detectors \textit{D}$_{1,2}$
shown in Fig. \ref{oal-setup}.  Each photoelectric pulse from
\textit{D}$_{1,2}$\ is stamped with its time of detection ($1$ ns
resolution) and then stored for later analysis, with examples of
the record of output counts versus time displayed in Fig.
\ref{rvst}.

\textbf{Acknowledgements} We gratefully acknowledge interactions
with K. Birnbaum, C.-W. Chou, A. C. Doherty, L.-M. Duan, T. Lynn,
T. Northup, S. Polyakov, and D. M. Stamper-Kurn. This work was
supported by the National Science Foundation, by the Caltech MURI
Center for Quantum Networks, and by the Office of Naval Research.

\textbf{Competing interests statement} The authors declare that they have no
competing financial interests.

\textbf{Correspondence} and requests for materials should be
addressed to H.J.K. (e-mail: hjkimble@caltech.edu).


\begin{thebibliography}{99}
\bibitem{mu92} Mu, Y. \& Savage, C. M. One-atom lasers. \textit{Phys. Rev. A} \textbf{46}, 5944-5954 (1992).

\bibitem{ginzel93} Ginzel, C., Briegel, H.-J., Martini, U., Englert, B.-G. \& Schenzle, A.
Quantum optical master equations: The one-atom laser.
\textit{Phys. Rev. A} \textbf{48}, 732-738 (1993).

\bibitem{pellizzari94a} Pellizzari, T. \& Ritsch, H. Preparation of
stationary Fock states in a one-atom Raman laser. \textit{Phys.
Rev. Lett.} \textbf{72}, 3973-3976 (1994).

\bibitem{pellizzari94b} Pellizzari, T. \& Ritsch, H. Photon statistics of
the three-level one-atom laser. \textit{J. Mod. Opt.} \textbf{41},
609-623 (1994).

\bibitem{horak95} Horak, P., Gheri, K. M. \& Ritsch, H. Quantum dynamics of
a single-atom cascade laser. \textit{Phys. Rev. A} \textbf{51},
3257-3266 (1995).

\bibitem{meyer97a} Meyer, G. M., Briegel, H.-J. \& Walther, H. Ion-trap
laser. \textit{Europhys. Lett.} \textbf{37}, 317-322 (1997).

\bibitem{loeffler97} L\"{o}ffler, M., Meyer, G. M., \& Walther, H. Spectral
properties of the one-atom laser. \textit{Phys. Rev. A}
\textbf{55}, 3923-3930 (1997).

\bibitem{meyer97b} Meyer, G. M., L\"{o}ffler, M. \& Walther, H. Spectrum of
the ion-trap laser. \textit{Phys. Rev. A} \textbf{56}, R1099-R1102
(1997).

\bibitem{meyer98} Meyer, G. M. \& Briegel, H.-J. Pump-operator treatment of
the ion-trap laser. \textit{Phys. Rev. A} \textbf{58}, 3210-3220
(1998).

\bibitem{jones99} Jones, B., Ghose, S., Clemens, J. P., Rice, P. R. \&
Pedrotti, L. M. Photon statistics of a single atom laser.
\textit{Phys. Rev. A} \textbf{60}, 3267-3275 (1999).

\bibitem{kilin02} Kilin, S. Ya. \& Karlovich, T. B. Single-atom laser:
Coherent and nonclassical effects in the regime of a strong
atom-field correlation. \textit{JETP} \textbf{95}, 805-819 (2002).

\bibitem{mckeever03} McKeever, J. \textit{et al}. State-Insensitive
Cooling and Trapping of Single Atoms in an Optical Cavity.
\textit{Phys. Rev. Lett.} \textbf{90}, 133602 (2003).

\bibitem{sargent-book} Sargent III, M., Scully, M. O., \& Lamb,
Jr., W. E. \textit{Laser Physics} (Addison-Wesley, Reading Mass.,
1974).

\bibitem{haken-book} Haken, H. \textit{Laser Theory} (Springer Verlag,
Berlin, 1984).

\bibitem{mandel-wolf-95} Mandel, L. \& Wolf, E. \textit{Optical Coherence and Quantum Optics}
(Cambridge Univ. Press, Cambridge, 1995).

\bibitem{carmichael-book} Carmichael, H. J. \textit{Statistical Methods in Quantum Optics 1}
(Springer-Verlag, Berlin, 1999).

\bibitem{gardiner-book} Gardiner, C. W. \& Zoller, P. \textit{Quantum Noise}
(Springer-Verlag, Berlin, 2000).

\bibitem{hjk sweden} Kimble, H. J. Strong interactions of single atoms and
photons in cavity QED \textit{Phys. Scr.} \textbf{T76}, 127
(1998).

\bibitem{walther-review} Raithel, G., Wagner, C., Walther, H., Narducci, L. M.
\& Scully, M. O. in \textit{Cavity Quantum Electrodynamics} (ed.
Berman, P.) 57-121 (Academic Press, San Diego, 1994).

\bibitem{haroche-review} Haroche, S. \& Raimond, J. M. in \textit{Cavity Quantum
Electrodynamics} (ed. Berman, P.) 123-170 (Academic Press, San
Diego, 1994).

\bibitem{meystre92} Meystre, P. in \textit{Progress in Optics, Vol. XXX}
(ed. Wolf, E.) 261-355 (Elsevier Science Publishers B.V.,
Amsterdam, 1992).

\bibitem{feld-ml} An, K. \& Feld, M. S. Semiclassical four-level
single-atom laser. \textit{Phys. Rev. A} \textbf{56}, 1662-1665
(1997).

\bibitem{chang-campillo96} Chang, R. K. \& Campillo, A. J. (eds) \textit{Optical Processes in Microcavities}
(World Scientific, Singapore, 1996).

\bibitem{vahala03} K. J. Vahala, Optical microcavities, \textit{Nature} (2003).

\bibitem{ye99} Ye, J., Vernooy, D. W. \& Kimble, H. J. Trapping of
single atoms in cavity QED. \textit{Phys. Rev. Lett.} \textbf{83},
4987-4990 (1999).

\bibitem{rice94} Rice, P. R. \& Carmichael, H. J. Photon statistics of a
cavity-QED laser: A comment on the laser-phase-transition analogy.
\textit{Phys. Rev. A} \textbf{50}, 4318-4329 (1994).

\bibitem{boozer03} Boozer, A. D., Boca, A., Buck, J. R., McKeever, J. and
Kimble, H. J., Comparison of Theory and Experiment for a One-Atom
Laser in a Regime of Strong Coupling, Phys. Rev. A (submitted,
2003); preprint available at
\textit{http://lanl.arxiv.org/archive/quant-ph}.

\bibitem{metcalf99} Metcalf, H. J. \& van der Straten, P. \textit{Laser Cooling and Trapping}
(Springer-Verlag, 1999).

\bibitem{boiron96} Boiron, D. \textit{et al}. Laser cooling of cesium atoms
in gray optical molasses down to 1.1 $\mu$K. \textit{Phys. Rev. A}
\textbf{53}, R3734-R3737 (1996) and references therein.

\bibitem{corwin99} Corwin, K. L., Kuppens, S. J. M., Cho, D. \& Wieman, C. E.
Spin-polarized atoms in a circularly polarized optical dipole
trap. \textit{Phys. Rev. Lett.} \textbf{83}, 1311-1314 (1999).
\end{thebibliography}
\end{document}